\documentclass{eas}
\usepackage{graphicx}
\providecommand{\figref}[1]{Fig.~\ref{#1}}
\providecommand{\secref}[1]{Sect.~\ref{#1}}
\begin{document}

\title{Study of short period variables and small amplitude periodic variables} 
\runningtitle{Varadi \etal: Study of short period variables \dots}
\author{Mihaly Varadi}\address{Geneva Observatory, University of Geneva, ch. des Maillettes 51, CH-1290 Sauverny, Switzerland}
\author{Laurent Eyer}\sameaddress{1}
\author{Stefan Jordan}\address{ARI/ZAH, Univ. of Heidelberg, M\"{o}nchhofstr. 12-14, D-69120 Heidelberg, Germany}
\author{Detlev Koester}\address{Institut f\"{u}r Theoretische Physik und Astrophysik, University of Kiel, Leibnizstra\ss e 15, D-24098 Kiel, Germany}
\begin{abstract}
Our goal is to assess Gaia`s performance on the period recovery of short period ($p < 2$ hours) and small amplitude variability. To reach this goal first we collected the properties of variable stars that fit the requirements described above. Then we built a database of synthetic light-curves with short period and low amplitude variability with time sampling that follows the Gaia nominal scanning law and with noise level corresponding to the expected photometric precision of Gaia. Finally we performed period search on the synthetic light-curves to obtain period recovery statistics. This work extends our previous period recovery studies to short period variable stars which have non-stationary Fourier spectra.
\end{abstract}
\maketitle
\section{Introduction \label{sect:intro}}

Variability is a common phenomena of stars. In the last decades the large photometric surveys observed different fraction of variability, probably because these surveys had different photometric precision, time resolution and time sampling. Gaia expected to detect 5 - 15 percent of variability (Eyer \& Cuypers \cite{eyercuypers2000}) This means that around $\sim$100 million variable star light curves will have to be processed by the Gaia DPAC CU7.  To have a glimpse about how Gaia will deal with variable see Eyer \etal  in these proceedings. 

Studying variable stars contribute to many field of astrophysics, just to mention a few:
With the methods of asteroseismology one can look inside the stars accessing information of the stellar interior which let one to probe stellar evolution and stellar pulsation theories. With the famous period-luminosity relations of classical Cepheids and RR Lyrae  stars one can determine the distance scales inside the Milky Way and to nearby galaxies. Studying compact pulsators like ZZ Ceti stars one can explore the physics of degenerate matter a regime which is not or hardly accessible in the laboratories. 

Our study is focusing on stars which shows variability in timescale shorter than 2 hours, hereafter short period variables. We present their types and properties in \secref{sect:obj} . These kind of variable stars, once detected, are advantageous from observational point of view as in a few hours long observation run many variation cycle is possible to be followed-up.  
In the scope of Gaia satellite, we are interested to know that what kind of information can be obtained from the observed Gaia light-curves of short period variables. For example detection of short period variability is one of the questions we are interested in, the possibility to recover their periods is another one. In this proceedings we are exploring the latter case, by analysing the Fourier spectra of simulated Gaia light-curves of short period variable stars. We choose the type ZZ Ceti as basis for our analysis, because observed ZZ Ceti star light-curves show a few level of complexity which represents mostly the of observed behaviours of short period variable stars. We are discussing such observed complexities  together with the types and properties in \secref{sect:obj}. In the \secref{sect:mission} we describe the mission properties of Gaia which are relevant to our study, in the \secref{sect:datasim} we describing the simulation and in the \secref{sect:result} we presenting the period recovery results on simulated data. 

\section{Types and properties of short period variable stars\label{sect:obj}}

Many type of variable stars show variability in time scales shorter than 2 hours as shown in the Table \ref{typesandprops}.
Most of the types are non-radial pulsators so their variability amplitudes are in the millimagnitude level and they often shows multiperiodic variability. In some of the short period variable star types nonlinearities in the light-curves are commonly observed. For example in the case of ZZ Ceti stars two different attempts were made to theoretically explain and model such nonlinearities. One of them is a model developed by Brickhill (\cite{brickhill1990}) then later presented, explained and used by others (Wu \cite{wu1997}; Goldreich \& Wu \cite{goldwu1999}; Wu \cite{wu2001}; Ising \& Koester \cite{isingko2001}; Montogmery \cite{montgomery2005}) which includes interaction between the pulsation and convection zone, thus relates the origin of nonlinearities to the propagation of the flux trough the convective zone.
According to the  other model by Brassard \etal  (\cite {brassard1995}) the deviations from linearity in the emergent flux are caused by the radiation transfer process. These nonlinearities becomes larger if the amplitude of the temperature perturbation increases. This process is neglected in the theory of Brickhill, however models based on their calculation can be fitted well to the observed mono- and multiperiodic ZZ Ceti stars (Montgomery \cite{montgomery2005}; Montgomery \cite{montgomery2007}). 

Further complexities are also observed in the ligth-curves of ZZ Ceti stars. For example Handler \etal  \cite{handler2008} reported
amplitude and frequency variation of the modes of the ZZ Ceti star EC 14012-1446. The cause of the amplitude changes on timescales from weeks to years of white dwarf variables is not well understood (see Winget \& Kepler \cite{winget2008}).

\begin{table}[htdp]
\caption{Types and properties of short period variables}
\begin{center}
\begin{tabular}{|c|c|c|}
\hline
\bf{Type} & \bf{Periods [minutes]} & \bf{Amplitudes [mag]} \\ \hline
$\beta$ Cep stars&96 - 480 &$<0.1$  \\ \hline
$\delta$ Scuti stars&28 - 480 &0.003 - 0.9  \\ \hline
roAp stars& 6 - 21 & $< 0.01$ \\ \hline
EC14026 stars&1.3 - 8.3  &$ <0.03$ \\ \hline
Betsy stars (PG1716)&33 - 150 & $<0.03$ \\ \hline
ZZ Ceti stars (DAV)&0.5 - 25 &$0.001 - 0.3$ \\ \hline
V777 Her stars (DBV)&2 - 16 &$0.001 - 0.2$ \\ \hline
GW Vir stars (DOV + PNNVs)&5 - 85&0.001-0.2 \\ \hline
Brown Dwarf pulsators&~60 - ~120 & -  \\ \hline
eclipsing white dwarfs & $>6$ & $<0.75$ \\ \hline
\end{tabular}
\end{center}
\label{typesandprops}
\end{table}%

\section{Relevant mission properties \label{sect:mission}}

\begin{figure} 
	\begin{center} 
		\includegraphics[width=9.0cm]{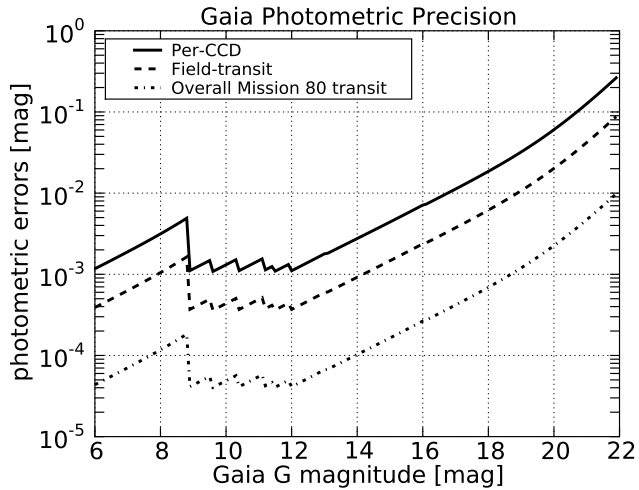} 
	\end{center}
	\caption{Expected photometric precisions of Gaia: per-CCD, field transit, and in the end of mission. Figure by M.Varadi using Gaia DPAC CU5 Photpipe.}
	\label{figure1}
\end{figure}

Gaia will observe about one billion stars with unprecedented astrometric and photometric precision (see \figref{figure1}) down to magnitude G ~ 20 mag. Over its 5 year mission, the two field of  view of Gaia will systematically scan all the sky according to the Gaia nominal scanning law and the observed sources will cross the CCDs of the common focal plane about 70 times on average. The detector in the focal plane consist 9 CCD in the astrometric field and each of them integrates charges for 4.4 seconds as the star crossing the CCD in TDI (Time Delay Integration) mode. This means 9 observation during one cross, and combining all the focal plane crosses a star on average will have about 630 CCD measurements in the G band. This is called G band per-CCD photometry, the averaged value of the 9 CCD measurements called Field of View (FoV) transit photometry, the averaged value over all the mission called end of mission photometry. 

The time gap between two consecutive field of view transit  is defined by the satellite design and by the Gaia nominal scanning law. This semi-regular time sampling defines the spectral window (SW) of a star on a given position on the sky. This spectral window usually has high values ($SW(f) \sim 0.6-0.9$) at higher frequencies ($f >= 12$ [c/d]) but sometimes a higher peak ($SW(f) \sim 0.6$ )appears around $4$ [c/d] which corresponds to the rotation of the satellite.  
 
The variability amplitudes of short period variables are mostly at millimagnitude level. As comparison the expected photometric precisions of Gaia is also at the millimagnitude level in wide range of G magnitudes as it is shown in \figref{figure1}. 

\section{Light-Curve simulations \label{sect:datasim}}
In order to simulate G band Gaia time series of short period variable stars, first we are using a light-curve model to generate continuous light-curves, then we sample the continuous model light-curve with the nominal scanning law of Gaia, in the end we add Gaussian noise which
corresponds to the expected photometric precision of Gaia (see \figref{figure1}).  We use the following models according to the required complexity level: mono-periodic, multiperiodic,  multiperiodic with nonlinearities (ZZ Ceti), multiperiodic with non stationary spectra. The inputs of the models are periods and amplitudes which corresponds to their observed values in real ZZ Ceti stars. 

\section{Analysis of period recovery statistics \label{sect:result}}

We derive the period recovery statistics from a sample of 200 light-curves. Each of them was generated with equal number of measurements, with different time sampling (these corresponds to stars with different sky positions) and with the same periods and amplitudes as inputs for the models.  Then we compute the Deeming periodogram (see Deeming \cite{deeming1975}) for each light-curve until a chosen maximum frequency value. After this we select the maximum peaks frequency and we compare with one of the input periods. If there is a match within the width of the peak then we count the period as successfully recovered. After pre-whitening we repeat the process if we are interested in the recovery of secondary frequencies.

\subsection{Mono-periodic model}
\begin{figure}
	\begin{center} 
	\includegraphics[width=9.0cm]{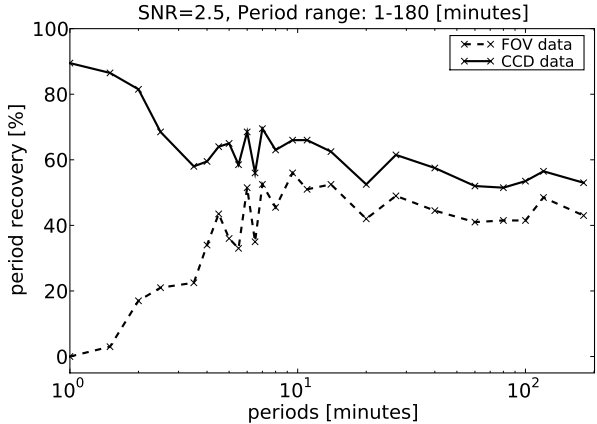} 
	\end{center}
	\caption{Period recovery percentages of mono-periodic signals in the short period regime. Solid lines corresponds to per-CCD photometry, Dashed lines corresponds to Field of View transit photometry. The signal to noise ratio is equal to 2.5 on the per-CCD measurements.}
	\label{figure2}
	
\end{figure}

Eyer and Mignard \cite{eyermignard2005} made an extensive research on the correct detection rate of mono-periodic signals of Gaia Field of View transit time series for a wide range of periods. They concluded that periods of regular variable star can be recovered even from signals with low S/N ratio and that the period recovery depends mainly on the ecliptic latitude. In the \figref{figure2} we present a result which extends their work to the short period regime and compares the period recovery from per-CCD photometry with period recovery from Field of View transit photometry.
The figure shows that for periods smaller than 5 minutes the period recovery on the per-CCD photometry is significantly higher than the period recovery on Field of View transit photometry. This is because the duration of the transit starts to be comparable with the period of the signal and in the case of per-CCD photometry this transit duration which is now significant part of the period is well sampled while it is averaged to a single value in the case of Field of View photometry.

\subsection{Multiperiodic models}
The multiperiodic models consist three cases: The first model consists a sum of sinusoidal signals with periods and amplitudes corresponding to the observed periods and amplitudes of a ZZ ceti star. The second is based on the first but taking into account the nonlinearities
caused by the convection zone as we described in \secref{sect:obj}. The nonlinear light-curve simulator code is the same what we described in Varadi \etal  \cite{varadi2009}. The third model is composed from two models of the first kind, let us call them A and B.
The model A is computed on the first half of the dataset (the first 2.5 year per-CCD observations) with a given input periods and amplitudes and the model B is computed on the second part of the dataset with different input periods and amplitudes. The input periods and amplitudes for the models A and B were taken from two different observational epoch of the ZZ Ceti star analysed by Handler \etal  \cite{handler2008}. This way we simulated Gaia time series of a ZZ Ceti star with non-stationary spectra.

We already published our results on the period recovery percentages on for the cases with first and second model (see Varadi \etal  \cite{varadi2009} ). Now we extend it with period recovery percentages for the model with the highest level of complexity. 
We found that for a G = 15 magnitude star about 73.5 percent of the cases the the correct period can be recovered if we use model of the first kind. This percentages drops only 1.5 percent if we use the second kind of model but it drops down to 17.5 percent if we use simple model with non-stationary spectra as described by the third kind of model.

We conclude that the non-stationarity of the spectra which is a commonly observed feature of white dwarf variables can cause serious challenge to our period recovery attempts on white dwarf variables from Gaia time series. Fortunately subdwarf variables (the types EC 14026 stars and Betsy stars) are having more stable modes which keeps their spectra stationary many years. (Charpinet 2009 private communications). 

\section{Aknowledgement}
This research project has been supported by ELSA under FP6 contract MRTN-CT-2006-033481.
We would like to say thank to the conference organisers for the beautifully organised conference.

\end{document}